\newif\iffigs\figstrue
\documentclass[12pt]{article}
  \input{epsf}
\else
\fi

\newcommand{\sect}[1]{\setcounter{equation}{0}\section{#1}}

\usepackage{amsmath}
\usepackage{amscd}
\usepackage{array}
\usepackage{latexsym,amssymb}

\begin{document}


\newcommand{\nn}{\nonumber}
\newcommand{\mm}{\hfill\cr}
\newcommand{\noi}{\noindent}
\def\di{\displaystyle}
\def\sx{\left}
\def\dx{\right}
\def\to{\rightarrow}
\def\ul{\underline}

\newcommand{\bm}[1]{\mbox{\boldmath $#1$}}

\newcommand{\be}{\begin{equation}}
\newcommand{\ee}{\end{equation}}
\newcommand{\bea}{\begin{eqnarray}}
\newcommand{\eea}{\end{eqnarray}}
\newcommand{\ov}{\overline}
\newcommand{\ba}{\begin{eqnarray}}
\newcommand{\ea}{\end{eqnarray}}

\def\sk{\vskip .4cm}

\def\a{\alpha}
\def\ap{\alpha'}
\def\b{\beta}
\def\c{\chi}
\def\cb{\ol {\chi}}
\def\d{\delta}
\def\e{\epsilon}
\def\f{\varphi}
\def\g{\gamma}
\def\k{\kappa}
\def\l{\lambda}
\def\m{\mu}
\def\n{\nu}
\def\o{\omega}
\def\p{\psi}
\def\r{\rho}
\def\s{\sigma}
\def\t{\tau}
\def\th{\theta}
\def\ve{\varepsilon}
\def\z{\zeta}

\def\D{\Delta}
\def\G{\Gamma}
\def\L{\Lambda}
\def\O{\Omega}
\def\S{\Sigma}
\def\Th{\Theta}


\newcommand{\rtr}{\mathrm{tr}}
\newcommand{\rU}{\mathrm{U}}
\newcommand{\rUSp}{\mathrm{USp}}
\newcommand{\rSU}{\mathrm{SU}}
\newcommand{\rE}{\mathrm{E}}
\newcommand{\rSO}{\mathrm{SO}}
\newcommand{\rSL}{\mathrm{SL}}
\newcommand{\cV}{\mathcal{V}}
\newcommand{\rSp}{\mathrm{Sp}}
\newcommand{\rF}{\mathrm{F}}
\newcommand{\rGL}{\mathrm{GL}}
\newcommand{\rG}{\mathrm{G}}
\newcommand{\rK}{\mathrm{K}}


\def\cA{\mathcal{A}}
\def\cB{\mathcal{B}}
\def\cC{\mathcal{C}}
\def\cD{\mathcal{D}}
\def\cE{\mathcal{E}}
\def\cF{\mathcal{F}}
\def\cG{\mathcal{G}}
\def\cH{\mathcal{H}}
\def\cI{\mathcal{I}}
\def\cJ{\mathcal{J}}
\def\cK{\mathcal{K}}
\def\cL{\mathcal{L}}
\def\cM{\mathcal{M}}
\def\cN{\mathcal{N}}
\def\cO{\mathcal{O}}
\def\cP{\mathcal{P}}
\def\cQ{\mathcal{Q}}
\def\cR{\mathcal{R}}
\def\cS{\mathcal{S}}
\def\U{\mathcal{U}}
\def\cV{\mathcal{V}}
\def\cW{\mathcal{W}}

\def\lag{{\mathcal{L}}}


\newcommand{\fgl}{\mathfrak{gl}}
\newcommand{\fu}{\mathfrak{u}}
\newcommand{\fsl}{\mathfrak{sl}}
\newcommand{\fsp}{\mathfrak{sp}}
\newcommand{\fusp}{\mathfrak{usp}}
\newcommand{\fsu}{\mathfrak{su}}
\newcommand{\fp}{\mathfrak{p}}
\newcommand{\fso}{\mathfrak{so}}
\newcommand{\fl}{\mathfrak{l}}
\newcommand{\fg}{\mathfrak{g}}
\newcommand{\fr}{\mathfrak{r}}
\newcommand{\fe}{\mathfrak{e}}
\newcommand{\ft}{\mathfrak{t}}


\def\tI{{\Lambda}}
\def\tJ{{\Sigma}}
\def\tK{{\Gamma}}
\def\tL{{\Delta}}
\def\tM{{\tilde M}}
\def\tN{{\tilde N}}
\def\dt{{\tilde d}}
\def\Dt{{\tilde D}}



\def\real{{\rm Re}\hskip 1pt}
\def\imag{{\rm Im}\hskip 1pt}
\def\cc{{\rm c.c.}}
\def\re{{\rm Re}\mathcal{N}}
\def\im{{\rm Im}\mathcal{N}}
\def\ii{\mathrm{i}}
\def\ib{{\ol {\imath}}}
\def\j{\jmath}
\def\jb{{\ol {\jmath}}}
\def\kb{{\ol  k}}
\def\lb{{\ol  \ell}}
\def\mb{{\ol  m}}
\def\nb{{\ol {n}}}
\def\rb{{\ol {r}}}
\def\sb{{\ol {s}}}


\def\trace{{\rm Tr}\hskip 1pt}
\def\diag{{\rm diag}}
\def\notin{\hbox{{$\in$}\kern-.51em\hbox{/}}}

\def\inbar{\vrule height1.5ex width.4pt depth0pt}
\def\IB{\relax{\rm I\kern-.18em B}}
\def\IC{\relax\,\hbox{$\inbar\kern-.3em{\rm C}$}}
\def\ID{\relax{\rm I\kern-.18em D}}
\def\IE{\relax{\rm I\kern-.18em E}}
\def\IF{\relax{\rm I\kern-.18em F}}
\def\IG{\relax\,\hbox{$\inbar\kern-.3em{\rm G}$}}
\def\IH{\relax{\rm I\kern-.18em H}}
\def\II{\relax{\rm I\kern-.17em I}}
\def\IK{\relax{\rm I\kern-.18em K}}
\def\IL{\relax{\rm I\kern-.18em L}}
\def\IN{\relax{\rm I\kern-.18em N}}
\def\IP{\relax{\rm I\kern-.18em P}}
\def\IQ{\relax\,\hbox{$\inbar\kern-.3em{\rm Q}$}}
\def\IR{\relax{\rm I\kern-.18em R}}
\def\IU{\relax\,\hbox{$\inbar\kern-.3em{\rm U}$}}
\def\ZZ{\relax\ifmmode\mathchoice{\hbox{\cmss Z\kern-.4em Z}}{\hbox{\cmss Z\kern-.4em Z}}{\lower.9pt\hbox{\cmsss Z\kern-.4em Z}} {\lower1.2pt\hbox{\cmsss Z\kern-.4em Z}}\else{\cmss Z\kern-.4em Z}\fi}
\def\IGam{\relax{{\rm I}\kern-.18em \Gamma}}

\newcommand{\iden}{\mbox{{1}\hspace{-.11cm}{l}}}
\def\bfnull{\relax{\rm O \kern-.635em 0}}


\def\de{{\rm d}}
\def\der{\partial}
\def\bos{{\rm bos}}
\def\na{\nabla}
\def\ol{\overline}
\def\ot{\otimes}
\def\imez{\frac{{\rm i}}{2}}
\def\mez{\frac{1}{2}}
\def\qu{\frac{1}{4}}
\def\we{\wedge}
\def\square{{\,\lower0.9pt\vbox{\hrule \hbox{\vrule height 0.2 cm \hskip 0.2 cm \vrule height 0.2 cm}\hrule}\,}}

\def\twomat#1#2#3#4{\left(\begin{array}{cc} \end{array} \right)}
\def\twovec#1#2{\left(\begin{array}{c} {#1}\\ {#2}\\ \end{array} \right)}

\def\thetabar{\bar \theta}
\def\al{\alpha}


\def\C{\mathbb{C}}
\def\Z{\mathbb{Z}}
\def\Hb{\mathbb{H}}
\def\Eb{{\bf E}}
\def\Rb{{\bf R}}
\def\Eb{{\bf E}}
\def\gb{{\bf g}}

\begin{titlepage}
\rightline{DISTA/PHYS-021/09} \rightline{January 2010} \vskip 2em
\begin{center}{\bf QUANTUM COMPUTING WITH SUPERQUBITS}
\\[3em]
{\bf Leonardo Castellani}, {\bf Pietro Antonio Grassi}, {\bf and
Luca Sommovigo}\\ [2em] {\sl Dipartimento di Scienze e Tecnologie
Avanzate and
\\ INFN Gruppo collegato di Alessandria,\\Universit\`a del Piemonte Orientale,\\ Via Teresa Michel 11, 15121
Alessandria, Italy}\\ [2.5em]
\end{center}

\begin{abstract}
\hspace{-.7cm} We analyze some aspects of quantum computing with super-qubits
(squbits). We propose the analogue of a superfield formalism, and
give a physical interpretation for the Grassmann coefficients in
the squbit expansion as fermionic creation operators of an auxiliary quantum system.
In the simplest case the squbit is a
superposition of one Bose $\otimes$ Bose  and one Fermi $\otimes$ Fermi state, and its norm is
invariant under a $U(2)$ group realized with Clifford-valued matrices.
This case can be generalized to a
superposition of $n_B$ bosonic and $n_F$ fermionic states,
with a norm invariant under $U(n_B+n_F)$. Entanglement between squbits, super quantum gates and
teleportation are discussed.

\end{abstract}

\vskip 7cm \noi \hrule \vskip.2cm \noi {\small
leonardo.castellani, pietro.grassi, luca.sommovigo@mfn.unipmn.it }

\end{titlepage}

\newpage
\setcounter{page}{1}

\sect{Introduction}

In recent work \cite{CKW,MW,Miyake} a measure for tripartite qubit
entanglement has been given in terms of Cayley's hyperdeterminant
\cite{Cayley}, a generalization of the usual determinant of a
square matrix to the case of cubic matrices. A supersymmetric
generalization of the hyperdeterminant was found in \cite{CGS} for
cubic supermatrices and inspired the construction of super-qubits
of ref. \cite{BDDR}.

Here we present a somewhat different formulation of super-qubits,
renamed {\sl squbits} for short. This formulation allows for a
physical interpretation of the expansion coefficients on the
super-Hilbert space basis. In order to construct a superposition
of bosonic and fermionic states one needs to circumvent the
superselection rules by introducing a second (auxiliary) quantum
system, containing bosons and fermions. Then we consider the
bosonic subspace in the tensor product of the two super-Hilbert
spaces, so that squbits have bosonic statistics. The Grassmann
coefficients of ref.\cite{BDDR} become here fermion creation
operators of the auxiliary system.

The plan of the paper is as follows: in Section 2 we introduce
squbits as a superfield expansion, and point out the difficulty
for a physical interpretation (and realization). An auxiliary
bose-fermi quantum system is then introduced, and Clifford squbits
are defined. Tensor products of these squbits are examined in
Section 3. Mixed states and their tensor products are the subject
of Section 4, where the properties of the corresponding density
matrix $\rho$ are studied. As for usual qubits, $\rho^2=\rho$ is
the condition for purity. Entanglement of squbits is treated in a
systematic way in Section 5. Unitary 1-squbit and 2-squbit
supergates are introduced in Section 6, including the
supersymmetry gate that exchanges bosonic and fermionic states.
In Section 7 we discuss super-teleportation, and observe
that some caution is necessary in using correlations in the
auxiliary quantum system between squbits that are separated in
space. Finally Section 8 contains conclusions and outlook.


\sect{Squbits}

\subsection{Grassmann squbits}

As in ref. \cite{BDDR}, we consider superpositions of bosonic and
fermionic states, with coefficients being Grassmann numbers rather
than complex numbers. We can choose from the start the orthonormal
basis of the super-Hilbert space  to contain an {\it equal number}
of bosonic and fermionic states, preparing the ground for
supersymmetry. In fact, using an expansion in the anticommuting
Grassmann coordinates $\theta_i$, leads to a  superfield
expansion of the squbit:

\begin{equation}\label{squbit}
|\psi \rangle = b | B \rangle + f_i \theta_i | F_i \rangle +
b_{ij} \theta_i \theta_j | B_{ij} \rangle + f_{ijk} \theta_i
\theta_j \theta_k  | F_{ijk} \rangle + \cdots
\end{equation}

\noi (sum on repeated indices, with $i < j < k \cdots$) where the
$|B \rangle$ states are bosonic and the $|F\rangle$ states are fermionic. All their
coefficients $b$ and $f$ are complex. For $N$ Grassmann
coordinates $\theta_i$, there are $2^{N-1}$ bosonic and $2^{N-1}$
fermionic states in the above expansion. Note that the denomination
   ``squbit'' is not really justified in this case, since $|\psi \rangle$ does
not reduce to the usual qubit when all $\theta$'s are set to zero. In forthcoming Sections
we will consider more general
squbits that indeed reduce to the usual qubits or qudits (states of a $d$-level
quantum system). All of these will be called squbits, even if their bosonic part
is a qudit.

The adjoint of $| \psi \rangle$ is:
\begin{equation}\label{squbitad}
 \langle \psi | = \langle B | b^* + \langle F_i| \thetabar_i f^*_i
 + \langle B_{ij} | \thetabar_j \thetabar_i b^*_{ij} +
 \langle F_{ijk} | \thetabar_k \thetabar_j \thetabar_i f^*_{ijk}
 + \cdots
 \end{equation}

 \noi where we have used the usual properties of conjugation of
 fermionic quantities (i.e. we do not use the superstar
 conjugation of ref. \cite{BDDR}).

 The usual norm of $| \psi \rangle$ is then a Grassmann number:
\begin{equation}
\langle \psi | \psi \rangle =
|b|^2 + |f_i|^2 \thetabar_i \theta_i + |b_{ij}|^2 \thetabar_j
\thetabar_i \theta_i \theta_j + |f_{ijk}|^2 \thetabar_k
\thetabar_j \thetabar_i \theta_i \theta_j \theta_k + \cdots
 \end{equation}

 By integration on the Grassmann coordinates with an appropriate
 weight, we can
 define a {\it positive} super-norm as :
\begin{equation}
\langle \langle \psi | \psi \rangle \rangle \equiv  \int e^{\sum_i \thetabar_i \theta_i} \langle
\psi | \psi \rangle~
\Pi_i d\thetabar_i d\theta_i = |b|^2 + \sum |f_i|^2  +
 \sum_{i<j} |b_{ij}|^2 + \sum_{i<j<k}|f_{ijk}|^2  + \cdots
 \end{equation}
\noi admitting the interpretation as a sum of probabilities.
Namely, the probability to ``find the system'' in one of the basis
states is just the square modulus of the corresponding complex
coefficient,  the state being normalized as $\langle \langle \psi | \psi \rangle \rangle =1 $. For a discussion on norms on super Hilbert spaces see for example ref. \cite{rudolph}.

Note that, even if we are able to give a probabilistic
interpretation of the coeffiecients in the squbit expansion, a
physical realization of the squbit remains problematic, see also
ref. \cite{BDDR}. The auxiliary anticommuting coordinates
$\theta_i$ are introduced in order that all terms in the
superposition in (\ref{squbit}) be bosonic, so that the superqubit
itself has Bose statistics. This is necessary to circumvent
superselection rules (see for ex. \cite{Haag}), which forbid a
superposition of bose states with fermi states.

To realize the superposition in (\ref{squbit}) as a quantum
mechanical construction, we need  a further quantum mechanical
system (with bosons and fermions) that we can tensor with the
original ($|B \rangle, |F\rangle$) system. Then we can restrict
our attention to the bosonic subsector of this tensor product.

In the next subsection we use dynamical $\theta_i$'s that are part
of a Clifford algebra. This provides, at least in principle,  a
way to a physical realization of squbits.


\subsection{Clifford squbits}

We assume that we have a quantum mechanical system whose Hilbert
states are either fermionic or bosonic. For example, we can
consider quantum electrodynamics with photons and electrons. We
denote by $|B_I\rangle$ the states for a single photon (we can
assume that $|B_I\rangle$ are the heliticity states) and we denote
by $|F_A\rangle$ the states for a single electron. They are
produced by acting with the field operators $A_I, \psi_A$ on the
vacuum of the Fock space. The Hilbert space ${\cal H}_{BF}$ of
physical states decomposes into bosonic and fermionic subspaces:
${\cal H}_{BF} = {\cal H}_B \oplus {\cal H}_F$. We would like to
combine those states in order to have superpositions of bosons and
fermions. For this we consider an auxiliary quantum mechanical
system described by the action $S = \int dt \, \bar\theta^i \dot
\theta_i$. This system has only zero mode states. The bosonic
states are $|0\rangle, \theta_i \theta_j |0\rangle, \dots$ and the
fermionic states are $\theta_i |0\rangle, \theta_i \theta_j
\theta_k|0\rangle,..$, where $|0\rangle$ is the Fock vacuum of the
system. The operators $\theta_i$, $\bar \theta^i$ satisfy the
Clifford algebra $\{ \bar\theta^{i}, \theta_j\} =\delta^i_{j}$,
$\{ \theta_{i}, \theta_j\} = 0$, $\{ \bar\theta^{i},
\bar\theta^j\} = 0$. The Hilbert space ${\cal H}^\theta$ for this
system is decomposed into two parts ${\cal H}^\theta_B \oplus
{\cal H}^\theta_F$ which again cannot be mixed because of the
superselection rules. Moreover, to tensor states of the ${\cal
H}^\theta$ system we follow the rule:
\begin{equation}
\theta_i \theta_j \cdots |0 \rangle \otimes \theta_k \theta_l \cdots |0 \rangle =
\theta_i \theta_j \cdots  \theta_k \theta_l \cdots |0 \rangle \label{tensortheta}
\end{equation}
\noi This rule is justified in the context of quantum field theory, where
for example the tensor product of two one-particle states is a two particle state
and so on.
\sk
Considering now the combined system with Hilbert space $ {\cal
H}_{BF} \otimes {\cal H}^\theta$ we can construct superpositions
involving bosons and fermions of the original ($|B \rangle,
|F\rangle$) system.  Supposing that there are $2^{N-1}$ bosons
$|B_I\rangle $ and $2^{N-1}$ fermions $|F_A\rangle$, we can
organize the states according to the binomial expansion
\begin{equation}
|\psi\rangle = b|0\rangle \otimes |B\rangle + f_i \theta_i
|0\rangle \otimes |F_i \rangle + b_{ij} \theta_i \theta_j
|0\rangle \otimes |B_{[ij]} \rangle + f_{ijk} \theta_i \theta_j
\theta_k |0\rangle \otimes |F_{[ijk]} \rangle + \dots \label{cliff}
\end{equation}
where we have rearranged the bosonic states $|B_I\rangle$ into the multiplet of bosonic states $(|B\rangle, |B_i\rangle, |B_{[ij]}\rangle, \dots)$ and the fermionic states into an analogous multiplet.
In this way all the terms of the expansion have the same (bosonic)
statistics, and we are considering only those vectors belonging to
the bosonic subspace of $ {\cal H}_{BF} \otimes {\cal H}^\theta$,
i.e. to the Hilbert space $${\cal H} = \Big({\cal H}_B \otimes
{\cal H}^\theta_B\Big) \oplus \Big({\cal H}_F \otimes {\cal
H}^\theta_F\Big)$$
This is a {\it bona fide} Hilbert space with norm induced by the norms
in ${\cal H}_{BF}$ and ${\cal H}^\theta$:
\begin{equation}
\langle \psi | \psi \rangle~
  = |b|^2 + \sum |f_i|^2  +
 \sum_{i<j} |b_{ij}|^2 + \sum_{i<j<k}|f_{ijk}|^2  + \cdots
 \end{equation}
as can be checked by using
\begin{equation}\label{cliffad}
 \langle \psi | = \langle B | \otimes  \langle 0 |b^* + \langle F_i|
 \otimes  \langle 0 | \thetabar_i f^*_i
 + \langle B_{ij} | \otimes  \langle 0 |\thetabar_j \thetabar_i b^*_{ij} +
 \langle F_{ijk} | \otimes  \langle 0 |\thetabar_k \thetabar_j \thetabar_i f^*_{ijk}
 + \cdots
 \end{equation}
 and the Clifford anticommutation rules.

 Thus the auxiliary
$\theta$ system allows
the dynamical mixing between fermions and bosons of the
${\cal H}_{BF}$ system, with a well-defined positive norm,
necessary for a probabilistic interpretation.

 The simplest squbit is constructed using just
two states $|B\rangle$ and $|F\rangle$ and introducing a single
$\theta$. The generic superstate is then
\begin{equation}\label{statA}
| \psi\rangle =  b |0\rangle \otimes |B\rangle + f \theta
|0\rangle \otimes |F \rangle
\end{equation}
where $b,f$ are the complex amplitudes. The positive-definite norm
of the state is $|b|^2 + |f|^2$. The Clifford coordinate $\theta$
, in contrast with the usual superspace coordinate $\theta$,  is a
dynamical quantity and we can act upon it with quantum operations.


\sect{Tensor Products}

Let us consider the tensor product of the simplest squbit with
itself. Imagining to realize squbits with identical particles,
we consider the symmetrized tensor product
${1 \over 2}(| \psi\rangle \otimes | \psi'\rangle + | \psi'\rangle \otimes |
\psi\rangle)$ (since $| \psi\rangle$ is a bosonic state):

$$
[ | \psi\rangle \otimes | \psi'\rangle ]_{symm}= [( b |0\rangle \otimes |B\rangle +
f \theta |0\rangle \otimes |F \rangle) \otimes (b' |0\rangle
\otimes |B\rangle + f' \theta |0\rangle \otimes |F
\rangle)]_{symm} = $$

\begin{equation}\label{tensA}
=bb' |00\rangle \otimes |BB\rangle + (b f' + b' f)
\theta|00\rangle \otimes |BF\rangle\,,
\end{equation}
where $|00 \rangle \equiv |0\rangle \otimes |0\rangle$,
$|BB\rangle \equiv |B \rangle \otimes |B \rangle$ and $|BF\rangle
\equiv {1 \over 2} (|B \rangle \otimes |F \rangle + |F \rangle
\otimes |B \rangle)$.
Moreover $\theta |0\rangle \otimes \theta |0\rangle =0$ because of
the rule (\ref{tensortheta}) and $\theta^2 =0$.

Identifying the vacuum of the Fock space $|0\rangle$  with the tensor product
of the vacuum of the single particle Hilbert spaces $|00\rangle$,
we can rewrite eq. (\ref{tensA}) as follows
\begin{equation}\label{tensB}
[| \psi\rangle \otimes | \psi'\rangle ]_{symm}= c |0\rangle \otimes
|BB\rangle + d\,  \theta |0\rangle \otimes |BF\rangle\,.
\end{equation}
where $c = bb' /\sqrt{|b|^2 |b'|^2 + |b f' + b' f|^2}$ and $d = (b
f' + b' f) /\sqrt{|b|^2 |b'|^2 + |b f' + b' f|^2}$. The tensor
product has produced again a bosonic state since we have tensored
the bosonic subspaces $({\cal H}_B \otimes {\cal H}_B^\theta)
\oplus ( {\cal H}_F \otimes {\cal H}_F^\theta)$ leading to the
Hilbert space $$\Big({\cal H}_B \otimes {\cal H}_B^\theta \otimes
{\cal H}_B \otimes {\cal H}_B^\theta\Big) \oplus \Big({\cal H}_B
\otimes {\cal H}_B^\theta \otimes {\cal H}_F \otimes {\cal
H}_F^\theta\Big)\,.$$
The tensor product of ${\cal H}_F \otimes {\cal H}_F^\theta$ with itself
does not contain any state since it would involve identical $\theta$ fermions
in the same physical states.
The absence of fermi-fermi parts in the tensor product
has important consequences, as discussed in
the forthcoming sections.

\sk
Symmetrized tensor products of copies of the simplest squbit
 remain in Hilbert subspaces of dimension 2, as we see in the
 example (\ref{tensB}). Thus the simplest squbit is not so
 interesting as element of a super quantum computer realized
 with identical particles, since in this case
 any number of these squbits leads always to a 2-state quantum
 system. However the situation changes if we tensor squbits which
 are not realised with identical particles: then the tensor
 product of two simplest squbits is
  (in the following we omit the $\theta$-vacuum $|0\rangle$)
  \begin{equation}
   | \psi\rangle \otimes | \psi'\rangle =
   bb' |BB\rangle + b f' \theta |BF \rangle  + b' f \theta|FB\rangle\
  \end{equation}
 where now $|BF \rangle = |B\rangle \otimes |F \rangle$ and
 $|FB \rangle = |F\rangle \otimes |B \rangle$ are not
  symmetrized states. This tensor product lives in a Hilbert
  space with one bosonic and two fermionic base states.
  \sk

 We can consider a more general squbit of
 the form:

\begin{equation}\label{newA}
|\psi\rangle =\sum_i^n b_i |B_i\rangle + \theta  \sum_i^n f_i
|F_i\rangle \,,
\end{equation}
with $n$ bosonic states $|B_i\rangle$ and $n$ fermionic
states $|F_i\rangle$, combined via a single
Clifford coordinate $\theta$. Its norm is
 $\sum_i^n \Big(|b_i|^2 +
|f_i|^2\Big)$. For identical particles the symmetrized tensor
product reads
\begin{equation}\label{newB}
(|\psi\rangle \otimes |\psi'\rangle)_{symm}  = \sum_{ij} b_i b'_j
|B_{(i} B_{j)}\rangle +\theta \, \sum_{ij} \Big(b_i f'_j + b'_i
f_j\Big) |B_i F_j\rangle
\end{equation}
with $n(n+1)/2 + n^2 $ states surviving in the tensor product. For
example when $n=2$ , 7 states enter the tensor product.
\sk

The $n=2$ case is the simplest generalization of the qubit, and
indeed its bosonic part is the usual 2-state qubit. Thus we will
call it the {\bf N=1 squbit}, since only one $\theta$ enters its
construction, and it reduces to the usual qubit when $f_i=0$.
Introducing more $\theta$'s leads to higher $N$
squbits. For example we can consider the $N=2$ squbit,
with two auxiliary coordinates $\theta_i$:
\begin{equation}\label{newC}
|\psi\rangle =\sum_i^2 b_i |B_i\rangle +   \sum_i^2 \theta_i f_i |F_i\rangle + \theta_1 \theta_2 |\widetilde B\rangle
\,,
\end{equation}
\noi i.e. a superqubit with 3 bosonic states $|B_i\rangle$ and
$|\widetilde B\rangle$ and 2 fermionic states $|F_i\rangle$.
Tensor products can be obtained as above by symmetrising
 $|\psi \rangle \otimes |\psi'\rangle$ in the case
  of identical particles.


\sect{Density matrix}

The density matrix for a pure state elementary squbit reads:

\begin{equation}
\rho = | \psi \rangle \langle \psi| = |b|^2 |B\rangle \langle B| + b  f ^*|B\rangle \langle F| \bar \theta
+ f b^* \theta |F\rangle \langle B| + |f|^2 \theta |F\rangle \langle F| \bar \theta
\end{equation}

\noi This is a hermitian matrix with $Tr \rho = |b|^2 + |f|^2 =1 $.
\sk
  Mixed states are linear combinations of squbits $| \psi_\al
  \rangle$:
  \begin{equation}
  |\psi_{mixed} \rangle = \sum_\al p_\al | \psi_\al \rangle
   \end{equation}
   where the classical probabilities $p_\al$ satisfy $\sum_\al p_\al =
   1$, and the squbits are normalized as $ \langle \psi_\al | \psi_\al \rangle=1$.
        The corresponding density operator
   \begin{equation}
    \rho_{mixed} = \sum_\al p_\al |\psi_\al \rangle \langle \psi_\al |
    \end{equation}
 has unit trace, and takes the form
\begin{equation}\label{purA}
\rho = a |B\rangle \langle B| + b |B\rangle \langle F| \theta^\dagger +
\bar b \, \theta |F\rangle \langle B| + c\,  \theta |F\rangle  \langle F| \theta^\dagger
\end{equation}
with $a,c \in \mathbb{R}$.  The trace normalization  $Tr{\rho} =1$
implies $c = (1-a)$.

The condition that $\rho$ represents a pure state can be expressed as
$\rho^2 = \rho$.  If we compute the square of the density matrix:
\begin{equation}\label{purB}
\rho^2 =
(a^2 + |b|^2) |B\rangle \langle B| +
b |B\rangle \langle F| \theta^\dagger +
\bar b \, \theta |F\rangle \langle B| + (|b|^2 + (1-a)^2)\,  \theta |F\rangle  \langle F| \theta^\dagger\,.
\end{equation}
we see that the purity condition amounts to  $a= a^2 + |b|^2$.  In this case
$\rho$ has eigenvalues $\lambda =1$ and $\lambda = 0$ corresponding to the
   orthogonal eigenvectors:
   \begin{equation}
   |\lambda=1\rangle =  b |B\rangle +  (1-a)  \theta |F \rangle,~~~
    |\lambda=0\rangle =  b |B\rangle  - a \theta |F \rangle
    \end{equation}

Let us now consider the tensor product of two mixed states. By
using the properties of the anticommuting $\theta$'s we obtain
\begin{eqnarray}\label{rhorhoprime}
\rho \otimes \rho' &=& a a'  |BB\rangle \langle BB|  + |BB\rangle (a
b' \langle BF| + ba' \langle FB|)  \theta^\dagger\nonumber\\
& +& \theta (a \bar b' |BF\rangle + \bar b a' |FB\rangle ) \langle BB|
\nonumber \\
&+ & \theta [ a(1-a') |BF \rangle \langle BF| + a'(1-a) |FB
\rangle \langle FB| \nonumber\\
&+& b \bar b' |BF \rangle \langle FB| + \bar b
b' |FB \rangle \langle BF|] \theta^\dagger
\end{eqnarray}
which is not normalized since $|FF\rangle$ states are absent in
the tensor product (because of $\theta^2=0$). Then the tensored
density matrix  must be normalized. After doing so, we can verify
that if $\rho$ and $\rho'$ describe pure states, then also $\rho
\otimes \rho'$ is a pure state, i.e. if  $\rho^2 = \rho$ and
${\rho'}^2 = \rho'$, we find  $(\rho \otimes \rho')^2 = (\rho
\otimes \rho') $.
\sk
For identical particles we must consider the symmetrized tensor
product
\begin{equation}\label{purC}
[\rho \otimes \rho' ]_{symm}=
a a'  |BB\rangle \langle BB| +
(a b' + a' b)  |BB\rangle \langle BF| \theta^\dagger
\end{equation}
$$ +(a' \bar b+ a \bar b')  \, \theta |BF\rangle \langle BB| +
\Big((1-a) a' + (1-a') a + b \bar b' + b' \bar b\Big) \,  \theta
|BF\rangle  \langle BF| \theta^\dagger\,. $$ with
$|BF\rangle
\equiv {1 \over 2} (|B \rangle \otimes |F \rangle + |F \rangle
\otimes |B \rangle)$.

When $\rho$ and
$\rho'$ describe pure states, the
symmetrized tensor product (even if we normalize it)
 is not a pure state, i.e.
$(\rho \otimes \rho')_{symm}^2 \neq (\rho
\otimes \rho')_{symm} $. Indeed the symmetrization
destroys the purity of the combined system, unless
$\rho=\rho'$. In fact, as discussed in next Section,
symmetrization of tensor products induces entanglement,
a well known fact since
the symmetrized state $|\psi \rangle \otimes
|\psi' \rangle + |\psi' \rangle \otimes |\psi \rangle $ is in
general entangled unless $|\psi\rangle  = |\psi'\rangle $.


\sect{Entanglement}

We
denote by $(n_B,n_F,N=1)$ a squbit of the form
\begin{equation}\label{enD}
|\psi \rangle = \sum_i^{n_B} b_i |B_i\rangle + \theta \sum_I^{n_F}
f_I |F_I\rangle\,,
\end{equation}
and study the tensor product
\begin{equation}   \label{prod}
|\psi \rangle \otimes |\psi' \rangle = b_ib_j' |B_iB_j\rangle
+ b_i f_J' |B_i F_J \rangle + f_I b_j' |F_I B_j \rangle
\end{equation}

A generic state $|{\tilde \Psi}\rangle$ living in the tensor
product of the $(n_B + n_F)$-dimensional Hilbert spaces spanned by
the basis vectors $|B_i\rangle, |F_I \rangle$ can be written as
\begin{equation} \label{tenstate}
|{\tilde \Psi}\rangle = \tilde b_{ij} |B_i B_j \rangle + \tilde f_{iJ} |B_iF_J\rangle
+ \tilde g_{Ij} |F_IB_j\rangle
\end{equation}
When can we write this state in the
form of a tensor product as in (\ref{prod})? Or in other words, when
do the equations
 \begin{equation} \label{factorequations1}
   b_ib_j'=\tilde b_{ij},~~~b_if_J'=\tilde f_{iJ},~~~
   f_Ib_j'=\tilde g_{Ij}
   \end{equation}
admit a solution for $b,f,b',f'$ in terms of $\tilde b, \tilde f, \tilde g$ ?

First we resort to
a simple counting argument. Each $(n_B,n_F,N=1)$ squbit depends
on $2n_B+2n_F-2$ real numbers (indeed one has to subtract
from $2n_B+2n_F$ real parameters one arbitrary overall phase and
the normalization condition $\langle \psi|\psi \rangle = 1$).
 Thus the tensor product in (\ref{prod}) depends on
 $2(2n_B+2n_F-2)$ real numbers.

 On the other hand the generic state (\ref{tenstate}) depends
 on $2n_B^2 + 4n_Bn_F-2$ real parameters. Thus when
 \begin{equation}
  2(2n_B+2n_F-2)  \ge  2n_B^2 + 4n_Bn_F-2 \label{ineq1}
  \end{equation}
  the number of parameters describing the states $|\psi \rangle$ and
  $|\psi'\rangle$ is sufficient to solve the equations
  (\ref{factorequations1}).  This happens only for
   $n_B=0$ (a trivial case, yielding vanishing tensor products)
    and $n_B =1$. For example,
   with $n_B=1$, $n_F=1$, the generic two-squbit state
\begin{equation} \label{tensqubit}
|{\tilde \Psi}\rangle = \tilde b |B B \rangle + \tilde f |BF\rangle
+ \tilde g |FB\rangle
\end{equation}
can be written as the product
\begin{equation}
 ( |B\rangle+ \theta ~ {\tilde g \over \tilde b}  |F\rangle)
 \otimes ( \tilde b |B\rangle +\theta  \tilde f |F \rangle)
 \end{equation}

\noindent (for arbitrary $n_F$ the same factorization holds with $\tilde g  |F\rangle \rightarrow
\sum_{I=1}^{n_F} \tilde g_I   |F_I \rangle$ and $\tilde f  |F\rangle \rightarrow
\sum_{I=1}^{n_F} \tilde f_I   |F_I \rangle$ ).
Thus no entanglement is possible with tensor
products of $N=1$, $n_B=1$ squbits.

The situation is different for
squbits with bigger $N$ or $n_B$. For example, the squbit
\begin{equation}\label{enC}
|\psi\rangle  =
a \, |BB\rangle +
b \,   \theta_1 \theta_2   |F_1 F_2\rangle \,.
\end{equation}
cannot be written in a factorized form. In general, the
entanglement can be evaluated by computing the partial density
matrices and finding their eigenvalues. For $n_B \geq 2$, to
require that a state living in the tensor product of Hilbert
spaces be factorizable amounts to require extra conditions on the
parameters $\tilde b, \tilde f, \tilde g$ (cf. the purely bosonic case,
where the equations $ b_ib_j'=\tilde b_{ij}$
   can be solved only if $rank( \tilde b_{ij}) =1$).

The first squbit (in order of
complication) that may produce entangled states by tensoring is
the $n_B=2$, $n_F = 1$ squbit already considered in \cite{BDDR}.
\sk
We can repeat the above analysis for the case of identical particles.
We have to symmetrize the state (\ref{prod}), but by doing
so we obtain a state that is always entangled, unless
$|\psi \rangle = |\psi'\rangle$. Therefore we consider
the tensor product of $|\psi\rangle$ with itself:
\begin{equation}   \label{prodsame}
|\psi \rangle \otimes |\psi \rangle =  b_ib_j |B_iB_j\rangle
+ 2 b_i f_J |B_i F_J \rangle
\end{equation}
where now $ |B_i B_j \rangle \equiv {1\over 2} ( |B_i\rangle
\otimes |B_j \rangle +|B_j \rangle \otimes |B_i \rangle )$ and $
|B_i F_J \rangle \equiv {1\over 2} ( |B_i\rangle \otimes |F_J
\rangle +|F_J\rangle \otimes |B_i \rangle )$.
This tensor product is specified by $2n_B+2n_F-2$ real
parameters.

On the other hand we can write the generic state living
 in the tensor product
for identical particles:
\begin{equation} \label{tenstatesymm}
|{\tilde \Psi}\rangle =
\tilde b_{ij} |B_i B_j \rangle +
 \tilde f_{iJ} |B_iF_J\rangle
\end{equation}
depending on $n_B(n_B+1)+2n_Bn_F-2$ real parameters. When the
inequality
\begin{equation}
2n_B+2n_F-2 \geq n_B(n_B+1)+2n_Bn_F-2 \label{ineq2}
\end{equation}
holds, the number of parameters in (\ref{prodsame})
is sufficient to solve the equations:
 \begin{equation}
 b_ib_j = \tilde b_{ij},~~~2 b_i f_J = \tilde f_{iJ}
 \end{equation}
 Again for $n_B=0$ and $n_B=1$ the inequality is satisfied, and
 the generic state in the symmetric tensor space is factorizable.
 For example the two-squbit state with $n_B=1$, $n_F=1$:
  \begin{equation}
  |{\tilde \Psi}\rangle =
\tilde b |B B \rangle + \tilde f |BF\rangle
 \end{equation}
 can be factorized as
\begin{equation}
( \tilde b^{1\over 2} |B\rangle+ \theta ~{\tilde f \over 2 \tilde b^{1\over2}}
 |F\rangle)
 \otimes ( \tilde b^{1\over 2} |B\rangle+ \theta ~{\tilde f \over 2 \tilde b^{1\over2}}
 |F\rangle)
\end{equation}
\noindent (for arbitrary $n_F$ the same factorization holds with $\tilde f  |F\rangle \rightarrow
\sum_{I=1}^{n_F} \tilde f_I   |F_I \rangle$).
Thus we find that no entanglement is possible using
multiple squbit states for identical particles. Here too the
situation changes when $n_B \geq 2$ or when $N > 1$.


\sect{Supergates}

We examine here  the structure of linear operations
on the $N=1$ squbits in (\ref{newA}) that preserve their bosonic overall character and their norm, i.e. that
send physical states into physical states.

The most general linear operator $U$ preserving the bosonic character of the $N=1$ squbit
has the form:

\begin{eqnarray}
& & U=x_{ij} |B_i \rangle \langle B_j|
+ y_{ij} \theta |B_i \rangle \langle B_j|  \bar \theta +  r_{ij} \theta
|B_i \rangle \langle F_j|+ s_{ij} |B_i \rangle \langle F_j| \bar \theta
+ \nonumber \\
& &~~~~~~u_{ij}  \theta |F_i \rangle \langle B_j| +  v_{ij}  |F_i \rangle \langle B_j| \bar \theta
+ w_{ij}  |F_i \rangle \langle F_j| +  z_{ij}  \theta |F_i \rangle \langle F_j| \bar \theta
 \end{eqnarray}

\noi with $i,j=1,...n$. However four terms yield a null result on the squbit,
and therefore we can limit ourselves to consider operators of the form:

\begin{equation}
U=x_{ij} |B_i \rangle \langle B_j|
+ s_{ij} |B_i \rangle \langle F_j| \bar \theta
+  u_{ij}  \theta |F_i \rangle \langle B_j| +   z_{ij}  \theta |F_i \rangle \langle F_j| \bar \theta
 \end{equation}

It can be checked that the product of two such operators reproduces an operator
of the same form. In order to preserve the norm of the squbit this operator must be
unitary. Imposing  $U^\dagger U =I$ is equivalent to require the unitarity of  the $2n \times 2n$ matrix:
\begin{equation}
M= \left(
\begin{array}{cc}
x & s
\\
u & z
\end{array}\right)
\end{equation}
\noi Thus the supergate operator, not surprisingly,  depends
on the entries of the $U(2n)$ unitary matrix $M$, i.e. $U = U(M)$, and
the map $M\rightarrow U(M)$ preserves the product:
$U(M) U(M') = U(MM')$.
\noi In other words,  the supergate operators $U$ are representations (on the space of squbits)
of the group $U(2n)$.

We give now two examples of 1-squbit
supergate operators. For simplicity we consider $N=1, n_B=1, n_F=1$
(the simplest squbit), but the supergates can be immediately generalized
to higher values of  $n_B,n_F$.

\sk
\noi  {\bf Super-Hadamard gate}
\sk

\begin{equation}\label{op`E}
U_H = \frac{1}{\sqrt{2}} \Big( |B\rangle \langle B| +
|B\rangle \langle F |   \bar \theta +
\theta |F\rangle \langle B |
-  \theta |F\rangle \langle F| \bar \theta \Big)\,.
\end{equation}
\sk
\noi {\bf Supersymmetry gate}
\sk
It is in fact an $X$ gate for the $|B\rangle, |F\rangle$ system:
\begin{equation}
U_Q = |B\rangle \langle F| \bar \theta + \theta |F \rangle \langle B|
\end{equation}
\noi acting on the basis states as:
 \begin{equation}
  U_Q |B \rangle = \theta |F \rangle, ~~~U_Q \theta |F\rangle = |B\rangle~~~(but~U_Q |F\rangle =0)
   \end{equation}
   Note that $U_Q^2 = I$. A squbit  $|\psi\rangle= b|B\rangle + f \theta |F\rangle$ is {\it supersymmetric} if
   \begin{equation}
   U_Q |\psi\rangle =|\psi\rangle
   \end{equation}
   \noi   i.e. when $b=f$.

   Note that a supergate $U$ maps supersymmetric squbits into supersymmetric squbits
   if and only if it commutes with the supersymmetry gate. In this case $U$ has the form:
   \begin{equation}
   U=\rho {\mathrm{e}}^{\mathrm{i} \alpha}(|B\rangle \langle B| + |F\rangle \langle F| ) + {\mathrm{i}} \sqrt{1-\rho^2}~{\mathrm{e}}^{\mathrm{i}\alpha}  (|B\rangle \langle F| \bar \theta+
   \theta |F\rangle \langle B| )
   \end{equation}
 \noi with $\rho,\alpha \in R$.
   It is tempting to speculate that supersymmetry invariance could protect a supersymmetric
   state against decoherence (but then we would need a ``supersymmetric environment").


\subsection{CNOT gate}

Controlled quantum gates are important examples of multiple qubit
gates. Here we consider the simplest example of controlled supergate,
acting on two elementary squbits of the form $|\psi\rangle =b |B\rangle + f \theta |F\rangle$:
\begin{equation}
U_{CNOT} = \theta |BF \rangle \langle BB| + |BB \rangle \langle BF| \bar \theta
+ \theta |FB \rangle \langle FB| \bar \theta, \label{CNOT}
\end{equation}
\noi a hermitian and unitary operator, hence $U_{CNOT}^2=I$.
Recall that the tensor product space in this case has
dimension 3, with basis $|BB\rangle, |\theta |BF\rangle, \theta |FB\rangle$.
There is no $|FF \rangle$ basis element since $\theta |F\rangle \otimes \theta |F\rangle =0$.
Therefore the controlled supergate (\ref{CNOT}) can be represented by a $3 \times 3$
unitary and hermitian matrix, one dimension less than the usual bosonic CNOT gate.
If the first squbit (the control squbit) is in the state $\theta |F\rangle$, the $|B\rangle$ part of the second squbit
(the target squbit) remains unchanged, while the information of its $|F\rangle$ part vanishes because of
$\theta |F\rangle \otimes \theta |F\rangle =0$. If the first squbit is in the state $|B\rangle$,
the supersymmetry operator acts on the second squbit, exchanging $|B\rangle$ with
$\theta |F\rangle$. Note that in this gate the control squbit collapses on its
bosonic part.

Increasing $n_B$ and $n_F$ one  can consider more useful controlled supergates,
where the control squbit remains unchanged, and 1-squbit supergates are applied to
the target squbit.


\sect{Super-teleportation}

In this example of teleportation we will use $N=2$ squbits of the form
\begin{equation}
|\psi\rangle = b_0 |B_0\rangle + b_1 |B_1 \rangle + \theta_0 f_0 |F_0\rangle + \theta_1 f_1 |F_1\rangle
\end{equation}

Suppose that the squbit to be teleported between Alice and Bob is
a bosonic qubit $|\psi_C \rangle = a_0 |B_0\rangle + a_1 |B_1\rangle$.
Moreover, suppose that Alice and Bob have each one element of the
super-EPR pair:
$$\theta_0 \theta_1 ( |F_0 F_1\rangle - |F_1 F_0\rangle )/2$$
obtained from the singlet
in the tensor product of  two superqubits $|\psi\rangle$. Simple algebra yields:
\begin{equation}\label{teleA}
\Big(  a_0 |B_0\rangle + a_1 |B_1\rangle \Big) \otimes
\theta_0 \theta_1 ( |F_0 F_1\rangle - |F_1 F_0\rangle ) =
\end{equation}
$$
=
\theta_0 \theta_1 \Big( a_0 |B_0 F_0\rangle \otimes |F_1\rangle - a_0 |B_0 F_1\rangle \otimes |F_0\rangle + a_1 |B_1F_0\rangle \otimes |F_1\rangle - a_1 |B_1F_1\rangle \otimes |F_0\rangle
\Big)=
$$
$$
= \left[\frac{a_0}{a_1} \theta_0 |B_0 F_0\rangle + \theta_1 |B_0 F_1\rangle +
\theta_0  |B_1 F_0\rangle +  \frac{a_1}{a_0} \theta_1 |B_1 F_1\rangle\right]
\otimes \Big( a_0 \theta_0 |F_0\rangle + a_1 \theta_1 |F_1\rangle \Big)\,.
$$
The second bracket is the squbit of Bob: it is the susy transformed of the original state $|\psi_C\rangle$,
i.e. the state $U_Q |\psi_C \rangle$, where $U_Q$ is now the supergate that exchanges $|B_i\rangle$ with
$\theta_i |F_i\rangle$.

Then Bob applies $U_Q$ to this state, and recovers the original qubit $|\psi_C\rangle$, since
$U_Q^2=I$, {\it without need of any classical communication from Alice}. This clearly
violates causality, and is due to our assumption (\ref{tensortheta}) that implies $\theta_0 |F_0 \rangle \otimes
\theta_0 |F_0\rangle =0$ etc. This assumption can make sense only if the two factors
in the tensor product originate from the same region in space, as  for example
in a decay process. Here however we are assuming that the same holds for squbits that are
separated by a long distance: in this case the auxiliary quantum system we have
been using is inadequate to model squbit correlations, and indeed one should add
the $x$-dependent modes of the dynamical $\theta$'s. Then
$\theta_0(x) |F_0 \rangle \otimes  \theta_0 (x') |F_0\rangle$ does not vanish,
and classical communication becomes necessary for teleportation.

\sect{Conclusions and outlook}

In this Letter we provide a physical
 interpretation of superqubits (or more generally superqudits)  by introducing an auxiliary
 quantum mechanical system and combining it with the original
 Bose-Fermi system.  We have enlarged the class of superqubits considered in
\cite{BDDR}, and promoted their Grassmann
coefficients to Clifford dynamical variables,
which are fermionic creation operators of the auxiliary
quantum system.  As in
 \cite{BDDR} the resulting superqubit has bosonic statistics,
 and quantum supergates are now unitary matrices with Clifford entries,
 rather than supergroup matrices. It would be worthwhile to
 investigate their LOCC, SLOCC and entanglement classes (cf.
 \cite{BDDR,Borsten:2008wd} for Grassmann superqubits).

 Supersymmetry can be implemented, the supersymmetry operator
 being one of the supergates. If the environment would act
 on supersymmetric squbits only via supersymmetric quantum gates
 (and this is a  big if) then supersymmetry could provide a protection
 against decoherence.


\vfill \eject

\end{document}